\documentclass[manuscript]{aastex}

\usepackage{graphicx}
\usepackage{natbib}
\usepackage{color}
\fontsize{6.5pt}{0pt}\selectfont

\slugcomment{Not to appear in Nonlearned J., 45.}

\shortauthors{Oba et al.}

\begin{document}


\title{Height-dependent velocity structure of photospheric convection in granules and intergranular lanes with Hinode/SOT}


\author{T. Oba\altaffilmark{1,2}}
\affil{Department of Space and Astronautical science/SOKENDAI (The Graduate University for Advanced Studies)}
\email{oba.takayoshi@ac.jaxa.jp}


\author{Y. Iida\altaffilmark{3}}
\affil{Department of Science and Technology/Kwansei Gakuin University}
\author{T. Shimizu\altaffilmark{2}}
\affil{Institute of Space and Astronautical Science, Japan Aerospace Exploration
Agency}




\altaffiltext{1}{SOKENDAI (The Graduate University for Advanced Studies), 3-1-1 Yoshinodai, Chuo-ku, Sagamihara, Kanagawa 252-5210, Japan }
\altaffiltext{2}{Institute of Space and Astronautical Science, Japan Aerospace Exploration
Agency, 3-1-1 Yoshinodai, Chuo-ku, Sagamihara, Kanagawa 252-5210, Japan }
\altaffiltext{3}{Department of Science and Technology/Kwansei Gakuin University, Gakuen 2-1, Sanda, Hyogo 669-1337 Japan}

\begin{abstract}
　The solar photosphere is the visible surface of the Sun, where many bright granules, surrounded by narrow dark intergranular lanes, are observed everywhere. The granular pattern is a manifestation of convective motion at the photospheric level, but its velocity structure in the height direction is poorly understood observationally. Applying bisector analysis to a photospheric spectral line recorded by the \textit{Hinode} Solar Optical Telescope, we derived the velocity structure of the convective motion in granular regions and intergranular lanes separately. The amplitude of motion of the convective material decreases from 0.65 to 0.40 km/s as the material rises in granules, whereas the amplitude of motion increases from 0.30 to 0.50 km/s as it descends in intergranular lanes. 
These values are significantly larger than those obtained in previous studies using bisector analysis. 
The acceleration of descending materials with depth is not predicted from the convectively stable condition in a stratified atmosphere. Such convective instability can be developed more efficiently by radiative cooling and/or a gas pressure gradient, which can control the dynamical behavior of convective material in intergranular lanes. Our analysis demonstrated that bisector analysis is a useful method for investigating the long-term dynamic behavior of convective material when a large number of pixels is available. 
In addition, one example is the temporal evolution of granular fragmentation, in which downflowing material develops gradually from a higher layer downward. 
\end{abstract}





\section{Introduction}
\footnotesize{Solar granules are bright patterns surrounded by dark channels called intergranular lanes on the solar photosphere and are a manifestation of gas convection. Because convection is a dominant mechanism of energy transfer in the convection zone and photosphere, 
it is important to understand how granulation is created in the photosphere. 
Granulation is explained simply as follows: hot material rises in the granules, becomes cooler through radiative cooling, diverges horizontally, and is pulled down by gravity in intergranular lanes. Magnetohydrodynamic (MHD) numerical simulations have been used to reproduce granulation and have indicated the implications of the dynamics (\citealt{Stein1998}, \citealt{Hurlburt1984}).
They predicted that downflow structures are efficiently formed in intergranular lanes because of significant radiative loss of energy and the pressure gradient formed by granulation. Remarkable recent progress in the development of numerical simulations has made it possible to achieve a high spatial grid size on the order of 10 km. 
Using the MURaM simulation code \citep{Vogler2005}, \citet{Cheung2007} confirmed the importance of the radiative energy loss for producing photospheric granulation. Some observational works have examined the dynamical behavior of granules. They showed that upflows in the Dopplergrams coincide with bright granules in intensity images, whereas downflows are located in the intergranular lanes. The typical Doppler velocities are $\sim$1 km/s in both the granular and intergranular regions. 
To date, the spatial distribution and velocity field of granules have been vigorously investigated with ground-based visible-light observations (\citealt{Hirzberger1997}, \citealt{Berrilli1999}, \citealt{Roudier2003}). 
However, observations have not revealed the physical nature of granulation, in particular the detailed temporal behavior of granules, which are well simulated numerically. 
Considering the role of photospheric convection as an energy carrier along the height direction, the height variation of the convective velocity field, in addition to its spatial distribution, is important information for understanding the details of the convective dynamics. \\
　There are some difficulties in observationally investigating the vertical structure of the convective motions. 
It is especially important to spatially distinguish granules from intergranular lanes, which requires observations with subarcsecond spatial resolution. 
In any observations with insufficient spatial resolution, the Doppler velocities are significantly degraded by mixing of blue-shifted signals in granules with red-shifted ones in intergranular regions. 
We also need to properly remove the 5-min oscillations, which are global eigenmodes of compressive sound waves and cover the entire solar surface. The amplitude of the oscillations is typically equivalent to 0.3--0.4 km/s, which should not be considered negligible in studies of the convective Doppler signals \citep{Leighton1962}. \\
　There are two methods of deriving the velocity field at different heights. One is to use many absorption lines, each of which is formed at a different height. This method provided the typical velocity structure in the photosphere (\citealt{Durrant1979}, \citealt{Berrilli2002}). 
\citet{Kiefer2000} obtained the vertical root-mean-square (RMS) velocity from multiple lines and showed that both the upflows and downflows decrease with height. This method, however, does not provide uniform continuity with height, because the height coverage of different absorption lines may have large gaps. In addition, the absorption lines may not be recorded at the same time, depending on the instrumental configuration.\\
　Another method is bisector analysis of a single absorption line. Bisector analysis is based on the fact that the irradiance observed at each wavelength in the absorption line reflects the physical conditions at different heights. The profile around the line core is formed at a higher layer in the photosphere, whereas the irradiance in the line wings originates from a lower layer. One advantage of bisector analysis is that we can obtain the vertical velocity structures at exactly the same time with continuous height information, whereas the disadvantage is that it requires a sufficiently high spectral resolution and signal-to-noise (S/N) ratio to analyze the precise shape of the line profile, unlike multiple-line analysis. To achieve a high S/N ratio, previous works used spectral data recorded with large telescopes and integrated the time sequence data at the expense of the time resolution or summed over pixels. Using the Fe I 557.6 nm line observed with a ground-based telescope, \citet{Maltagliati2003} reported a difference in the velocity structure between granular and intergranular regions. They captured decelerating upward motion in granular regions, whereas the speed of the downflow was found to be constant over height in intergranular regions. \citet{Kostik2007} investigated the convective velocity structure over 570 km in granular regions and intergranular lanes using two lines, Fe II 523.4 and Fe I 639.3 nm, simultaneously recorded at the German Vacuum Tower telescope in Tenerife. They found no difference in the amplitude of typical velocities in the granular and intergranular regions; the speed of the upflows in granular regions decreases from 0.2 km/s to approximately 0 km/s over heights of 0 to 500 km, and the speed of the downflows shows the same height dependence as that in granular regions. \\
　A more sophisticated method of deriving the height gradient of the velocity in the atmosphere is Stokes inversion techniques, such as SIR (\citealt{RuizCobo1992}) and SPINOR (\citealt{Frutiger2000}). 
\citet{Frutiger2000} used SPINOR and \citet{Borrero2002} used SIR to derive the atmospheric height velocity profiles of granules and intergranular lanes from a disk-center spectral profile generated by integrating spatially and temporally. 
Both these authors reported that the upward speed in granules decreases from about 1 to 0 km/s at geometrical heights of 0 to 270 km; the intergranular lanes show velocity acceleration from 0 to 4 km/s in the same height range. 
The magnitude of the velocities in their studies is significantly larger than that derived using bisector analysis. 
 \citet{Rodriguez1999} applied SIR to spectral lines measured at each pixel and found that the upward velocity changes from roughly $<$1 km/s to 0.5 km/s at heights of 0 to 160 km in most of the data pixels for granules, which is roughly similar in magnitude to the velocity derived using bisector analysis. 
Among the studies described above, a fairly large deviation can be seen in the magnitude of the velocity as a function of height. There are two possible causes for this deviation: the velocity caused by 5-min oscillations and mixture of signals from granules and intergranular lanes. 
Most of the above studies ignored the effect of 5-min oscillations. 
The exception is \citet{Kostik2007}, in which a filtering process was performed to remove this effect.
However, the spectral data may be degraded under unstable conditions with atmospheric seeing. 
Our motivation is that previous works present a large deviation in their reported velocity magnitudes, while most of them did not perform a filtering process to remove the 5-min oscillations. In this study, therefore, we attempt to derive the convective height-structure accurately using a time series of stable, seeing-free high-resolution data with a filtering process to remove the 5-min oscillations.
The Solar Optical Telescope (SOT) \citep{Tsuneta2008} onboard \textit{Hinode} \citep{Kosugi2007} is suitable for this purpose.
SOT performs spectroscopic observations with diffraction-limited performance (0.3$^{\prime \prime}$ achievable with a 50-cm diameter) under stable conditions; thus, it has a remarkable advantage because its observations cover a much longer time than the typical period of 5 min for oscillatory motions and maintain a sufficiently high spatial resolution to clearly separate intergranular lanes from granules. In this study, we use a bisector method rather than inversion techniques because the simplest calculation for the velocity field can be used as the first step of analysis. 
Inversion techniques can also provide the line-of-sight velocity as a function of height, but they require skill and effort from users, who must provide an initial guess regarding an atmospheric model, fine-tune the free parameters, and remove strange results. \\
　In section 2, we describe the observations and data reduction. In section 3, we describe the bisector analysis and the method for removing the 5-min oscillations. Section 4 presents the results. The derived results are discussed in section 5. In section 6, we summarize our findings.

\section{Observations}
　The observations were made with the \textit{Hinode} SOT from 22:56 UT to 23:41 UT on 2014 July 6. The SOT has a primary mirror 50 cm in diameter, and its diffraction limit at 630 nm is about 0.3$^{\prime \prime}$, which corresponds to approximately 200 km on the solar surface. Diffractive imaging is achieved using the features of the developed optical system \citep{Suematsu2008} and real-time stabilization of images on the focal plane detectors \citep{Shimizu2008}. 
 In this study, we use the spectropolarimeter (SP) data \citep{Litesb2013} with blue continuum (450.45 nm, bandwidth 0.4 nm) images from the filtergraph (FG). The SP obtains all four Stokes profiles (I, Q, U, V) of two magnetic-sensitive Fe I spectral lines at 630.15 and 630.25 nm. The spectral resolution is 0.03 nm with CCD pixel sampling of 0.0215 nm. The series of SP data were acquired at a fixed slit position (``sit-and-stare'') near the disk center during the period. One SP exposure was recorded about every 2 s with an integrated exposure duration of 1.6 s, so 1,434 SP slit data were obtained (Fig. \ref{fig:intensity}). This high-cadence measurement allows us to trace the temporal evolution of the convective structures with a sufficient number of photons (S/N ratio higher than 900) for bisector analysis. The slit (width, 0.15$^{\prime \prime}$) is oriented in the solar N--S direction, and the N--S field of view (FOV) is 81.6$^{\prime \prime}$ with 0.16$^{\prime \prime}$ CCD pixels. The FG blue continuum images are used to monitor the evolution of the granular structures with time. One image was acquired every 30 s, so 90 images were produced during the period. The FOV is 19.2$^{\prime \prime}$ (EW) x 88.9$^{\prime \prime}$ (NS) with a pixel size of 0.11$^{\prime \prime}$. Because this study aims to investigate the pure convective motion in the absence of a magnetic field, the observed target is a quiet region where the averaged degree of polarization is less than 1\% in the entire FOV. Here the degree of polarization is defined as $V_{max}/I_{c}$, where $V_{max}$ is the highest Stokes V signal in the spectral profile. $I_{c}$ is the continuum intensity averaged over 0.01 nm at 630.1 nm in all the SP data. Because we focus on the convective structure in a quiet region, only Stokes I spectral data are used in the analysis. \\
　Bisector analysis is applied only to the Fe I 630.15 nm line because this line has two advantages compared to the other line. The first is that it has less magnetic sensitivity; the Fe I 630.15 nm line has a Land\'{e} factor of 1.5, whereas the Fe I 630.25 nm line has a Land\'{e} factor of 2.5. The second is that Fe I 630.15 nm is formed over a broader range than Fe I 630.25 nm. The contribution function of Fe I 630.15 nm (calculated by Prof. K. Ichimoto in 1995, private communication) shows that the line core intensity reflects a height of approximately 300 km above the $\tau = 1$ layer of the continuum at 500 nm, whereas Fe I 630.25 nm is formed at a height of 200 km. \\
　The SP data are calibrated using the standard routine SP\_PREP in the Solar SoftWare package (\citealt{Litesa2013}). The SP\_PREP routine performs the calibration, which includes i) dark-field correction, ii) flat-field correction, iii) compensation for residual Stokes I → Q, U, and V crosstalk, iv) removal of wavelength shifts on the period of the spacecraft orbit (about 98 min) caused by thermal deformation of the instrument optics, and v) calibration of intensity variations along the SP slit caused by tiny variations of the slit width. \\
　Regarding the absolute wavelength calibration of the spectral line for the reference velocity of 0 km/s, we utilized the wavelength calibrated by the SP\_PREP routine after confirming its validity. The mean line profile averaged spatially from our observations shows that the rest wavelength is slightly blue-shifted; i.e., it exhibits a convective blueshift. 
The magnitude of the convective blueshift depends on the formation height of the absorption lines \citep{Dravins1981}. 
\citet{Allende1998} investigated the rest wavelengths of 4947 absorption lines, including Fe I 630.15 nm, with high wavelength resolution.  According to their report, the convective blue shift of Fe I 630.15 nm is 0.21 km/s $\pm$ 0.11 km/s, which is in good agreement with the velocity of the mean line profile derived from our analysis (0.14 km/s).
The difference between \citet{Allende1998} and our analysis is 0.07 km/s, and we adopt 0.18 km/s as the worst error value in our results. 

\section{Analysis methods}
Two analysis methods are used to derive the pure convective velocity structure: bisector analysis, for deriving the Doppler velocity height structure, and an analysis to remove the 5-min oscillation signals from the spectral data. \\

\subsection{Bisector analysis} \label{bozomath}
　To derive the velocity field at different heights, we apply bisector analysis to the observed spectral line; in this method, the Doppler velocities at various intensity levels in the absorption line are calculated using the fact that each intensity in the absorption line reflects a different height. Because the absorption coefficient is maximum in the line core and decreases toward the wings, the line core and wings originate from higher and lower levels of the photosphere, respectively. Local thermodynamic equilibrium (LTE) is generally satisfied in the photosphere, meaning that the brightness is simply expressed as a certain temperature in a local region. 
In the optically thick regime, the source function increases linearly with height, meaning that the observed intensity can be determined from the intensity at around $\tau=1$ \citep{Stix2004}. 
Considering these two facts, an emergent intensity at a certain wavelength is given by only the temperature at $\tau=1$. 
We introduce a criterion to determine the range of intensity levels for calculating the velocity: We choose 0.10--0.15 of $I/I_{0}$ below the continuum intensity as the highest intensity level and an intensity of less than 0.05 of $I/I_{0}$ above the line core intensity as the lowest intensity level. We used a grid spacing of 0.05 for the intensity levels. Line profiles originating from granular regions typically have a higher continuum intensity, and the line cores (lower intensity levels) are deeper than those of intergranular lanes. An example of a line profile originating from granules is shown in Fig. \ref{fig:bisec} (a) and (b). The maximum intensity is approximately 1.13, and the minimum intensity is 0.26 in this case. The bisector is derived in the intensity range between 0.30 and 1.00 and consists of 15 intensity levels. On the other hand, the line profiles formed in intergranular lanes typically have a lower continuum intensity and higher intensity in the line core than those of granules, meaning that the number of intensity levels for the bisector is smaller than that for granular regions. An example is shown in Fig. \ref{fig:bisec} (c) and (d); the maximum intensity is approximately 0.92, and the minimum is 0.42. In this case, the bisector consists of 8 intensity levels covering values of 0.45 to 0.80. The statistical results are most reliable in the intensity range between 0.40 and 0.75 because the center of the absorption line cannot be calculated at higher or lower intensity levels in mainly intergranular regions. \\
　The bisector is converted to the Doppler velocity $v$ at each intensity level according to

\begin{equation}
v=c\frac{\Delta \lambda}{\lambda_{0}},
\label{eq:dop}
\end{equation}

\noindent where $c$ is the speed of light ($3.0 \times 10^{5}$ km/s), $\lambda_{0}$ is the wavelength of the absorption line ($630.15$ nm) without any motion, and $\Delta \lambda$ is the wavelength offset of the bisector from $\lambda_{0}$.\\

\subsection{Removal of 5-min oscillations} \label{bozomath}
　A subsonic filter \citep{Title1989} was applied to the data in the Fourier domain to extract the pure convective velocity from the measured velocity map. This process is composed of three steps. First, we created a $k-\omega$ diagram from the Doppler velocity field maps using the Fourier transformation. The Doppler velocity field map in our analysis has a spatial dimension (slit direction) and a temporal dimension, i.e., $y-t$ maps. Fig. \ref{fig:k?_sep} is an example $k_{y}-\omega$ diagram from the Fourier transformation. The Doppler velocity map at each intensity level is independently processed to derive the $k_{y}-\omega$ diagram. Second, we applied a subsonic filter to separate the pure convective motion and the 5-min oscillation signals. The signals in the phase velocity with $\omega/k_{y}$ $>$ 7 km/s, the sound speed in the photosphere, are regarded as the 5-min oscillation signals. This boundary is shown by the inclined dashed line in Fig. \ref{fig:k?_sep}. Below 1.5 mHz in the Fourier domain, we employed a different filter. 
The Fourier domain below 1.5 mHz and above $\omega/k_{y}$ $<$ 7 km/s is incorporated into the convective components because the minimum frequency of the p modes is considered to be 1.5 mHz. Third, the $k_{y}-\omega$ diagrams after filtering were transformed to $y-t$ space again, providing the pure convective motion and the 5-min oscillations separately. This process was applied to the time--distance diagram for each intensity level.\\

\section{Results}
\subsection{Separation of convective motion and 5-min oscillations} \label{bozomath}
　Fig. \ref{fig:td_sep} shows velocity diagrams before and after the filtering process at an intensity level of $I/I_{0}=0.70$. Panel (a) is a continuum intensity map, which is the $y-t$ image representing the morphological evolution of the granules at the slit position. 
Panel (b) is a velocity time--distance diagram before the filtering process. Panels (c) and (d) show the velocity diagrams of the pure convection and 5-min oscillations, respectively, separated by filtering. 
A comparison of panels (a) and (b) shows that the observed velocity is not correlated significantly with the continuum intensity. The thread-like patterns in panel (a) appear only in the velocity map after filtering. In panel (d), upward and downward speeds appear repeatedly with an approximately 5-min period. 


\subsection{{\bf Geometrical height}}
　It is important to determine the geometrical height to which each intensity level corresponds in the photosphere. 
The height of the formation layer is determined by the physical conditions along the path of light.
However, we do not have any way to derive most of the physical parameters from the spectral data.  
In this study, we focus on deriving a rough estimation of the geometrical height. 
We obtained geometrical heights simply using the average quiet Sun model in \citet{Vernazza1981}, as follows. 
Because the spectral line originates in the photosphere, in which LTE can be assumed, the Planck function can be adopted to relate the temperature at a certain height to the observed intensity at each wavelength in the spectral line.
Considering that the averaged temperature at 0 km is 6420 K in \citet{Vernazza1981}, we set the averaged continuum level of $I/I_{0}=1.0$ to a temperature of 6420 K. Intensity levels ranging from 0.40 to 0.75 at intervals of 0.05 are related to the temperature and thus the geometrical height according to the atmospheric model, as shown in Table 1. 
Note that the values given in Table 1 are typical geometrical heights. The intensity at a bisector level originates not only at a single geometrical height in the atmosphere but also at a fairly wide range of heights. Thus, the velocity derived by bisector analysis may be the weighted mean in a certain range of heights centered at the typical geometrical height.

\subsection{Relationship between convective velocity and continuum intensity}
　Fig.\ref{fig:td_cv} shows the relationship between the convective velocity after the 5-min oscillations are removed and the continuum intensity. Panel (a) shows the continuum intensity, and panels (b), (c), and (d) show the convective velocities at $I/I_{c}=0.75, 0.55$, and $0.40$, which are equivalent to heights of 40, 92, and 163 km, respectively. Note that a higher intensity level corresponds to a lower height. At the intensity level of $I/I_{c}=0.75$, i.e., the lowest layer closest to the continuum layer, blue and red patterns are clearly seen, indicating that there are strong convective velocities in both directions. 
As the intensity level goes to lower values, i.e., higher layers, the velocity contrast decreases with height. The velocity patterns seen in the velocity maps are quite similar to what is seen in the continuum map. \\
　Fig. \ref{fig:scp} shows scatter plots between the continuum intensity and the pure convective velocity at intensity levels of 0.75, 0.55, and 0.40. 
Note that Fig. 6 (a), (b), and (c) are scatter plots obtained using the velocity map before the filter was applied, whereas Fig. 6 (d), (e), and (f) are those obtained after the filtering process.
The correlation with the filtering process improves toward higher layers; the standard deviation of the distribution is 0.35, 0.27, and 0.22 km/s for intensity levels of 0.75, 0.55, and 0.40, respectively. The standard deviation is roughly 0.43 km/s in the data before filtering, showing that the filtering process enables us to obtain better correlations between the continuum intensity and convective velocity. \\

\subsection{Height dependence of velocities}
　Our method provides a time sequence of the convective velocity structures and 5-min oscillations in the vertical direction at a fixed slit position. An animation of the time evolution is available as supplemental material online. 
Fig. \ref{fig:movie} shows a snapshot of the time evolution of the velocity structures. 
Fig. \ref{fig:movie} (a) shows a two-dimensional (2D) spatial image of the blue continuum, which is scaled to the spatial resolution of the SP. Fig. \ref{fig:movie} (b) is a height--horizontal extent of the convective velocity along the white line in Fig. \ref{fig:movie} (a). Fig. \ref{fig:movie} (c) is the same as Fig. \ref{fig:movie} (b), but for the 5-min oscillations. \\
　Fig. \ref{fig:rms} shows the RMS velocities as a function of intensity level. 
In the unfiltered observed velocity plot, the RMS value of the Doppler velocity gradually decreases from 0.7 km/s at an intensity level of 0.75 to 0.5 km/s at an intensity level of 0.40. 
Similarly, the filtered convective velocity decreases from 0.6 km/s at an intensity level of 0.75 to 0.3 km/s at an intensity level of 0.40. This clearly shows that the magnitude of the convective motion is stronger in the lower layers and gradually decreases toward the higher layers. In contrast, the RMS amplitude of the 5-min oscillations increases from 0.3 km/s at an intensity level of 0.75 to 0.4 km/s at an intensity level of 0.40, indicating that the amplitude of the 5-min oscillations increases from the lower layers to the higher layers. The RMS amplitude of the convective motion is comparable to that of the 5-min oscillations at an intensity level of 0.45. At intensity levels higher than 0.45, the Doppler signals from the 5-min oscillations are less dominant than the magnitudes of the actual convective motion. At an intensity level of 0.75, the 5-min oscillatory velocities contribute approximately half the value of the convective velocity to the velocity field. \\
　We also investigated the averaged properties of upward and downward convective motion independently as a function of height (intensity level). Fig. \ref{fig:down_up} shows the averaged values of the upward and downward flows at each intensity level. The error bars give the standard deviation of the convective velocity at each intensity level. It is difficult to distinguish small velocities as upward or downward groups because of the uncertainty in the absolute wavelength. Because the estimated error of the velocity field is 0.18 km/s (see section 2), we defined the upward flows as the regions where the speed is lower than $-$0.18 km/s and the downward flows as those having speeds higher than 0.18 km/s.
Fig. \ref{fig:down_up} shows that the upward speed decelerates from 0.65 to 0.40 km/s with increasing height, whereas the downward speed accelerates from 0.30 to 0.50 km/s with increasing depth as the material moves into deeper layers of the photosphere.\\

\subsection{Granular fragmentation}
　Bisector analysis provides the details of the temporal evolution of the vertical velocity structures. Fig. \ref{fig:gra_frag} is an example showing the velocity--height structure at a slice located in a granule that fragmented during the measurement. 
The 2D morphology of the granule and its temporal evolution were captured in the FG blue continuum.
In the first frame, the granule is round. 
After 2 min (second row), the intensity starts to decrease in the center portion of the granule. In the third row, a high-intensity patch appears at the right edge of the granule, accompanied by a column showing upward velocity. The central portion of the granule shows a downward flow in a higher layer. The downward flow develops toward the lower layer (fourth row). The intensity continues to decrease, and the downward motion develops further (bottom row).
We found a total of four fragmentation events during the observation. 
The behavior described above was common in the other events. 
The scenario of the common temporal behavior is that a downflow signal appears in the upper layer and gradually develops toward a deeper layer, while the central portion of the granule simultaneously shows an intensity reduction. 
The downflow signal appears when the continuum intensity is reduced to around 1.0. Downflow from the upper layer to the lower layer (a distance of 160 km) develops on a short timescale, i.e., less than 30 s. Even after downflow is dominant over the entire height, the intensity continues to decrease, and finally a dark intergranular lane forms. \\

\section{Discussion}
\subsection{Convective structure}
In the previous section, we showed the average properties of convective flows as a function of height; the upward speed decreases from 0.65 to 0.40 km/s with increasing height, whereas the downward speed increases from 0.30 to 0.50 km/s with increasing depth. We also showed the temporal behavior of a fragmenting granule as an example to demonstrate that our bisector analysis is useful for investigating the long-term dynamical behavior of convective material when a large number of pixels is available.\\
　First, we discuss the validity of the subsonic filtering process for distinguishing convective motion from 5-min oscillation signals.
When a $k$--$\omega$ diagram is created from a long-duration series of images with a wide FOV, strong power ridges are clearly visible over 2--5 mHz, e.g., \citet{Stix2004}. In this study, we used a 45-min series of slit data that have only one spatial dimension with a short slit length, degrading the resolution of the wave number and time frequency in the $k$--$\omega$ diagram. 
In Fig. \ref{fig:k?_sep}, strong power is concentrated at 2--5 mHz, and a signal originating from convection is observed around 1 mHz, although discernible ridge signals are not clearly visible. This result means, therefore, that we can properly remove the 5-min oscillations by using the subsonic filter, even for such a time series of spectral data with one spatial dimension for a short slit length. \\
　High-quality spectral data from \textit{Hinode}/SOT were used in this study, which yielded results that differ quantitatively from those in previous works. 
For example, the convective velocities derived in this study are significantly larger than those in a previous work \citep{Kostik2007} in which bisector analysis was applied to data observed with a ground-based telescope. From Fig. \ref{fig:down_up}, the averaged velocity of the upward flows decreases from 0.65 to 0.25 km/s with increasing height, whereas their results show that the upflow speed in granular regions decreases from 0.2 to 0.1 km/s in the height range from 40 to 160 km, where the height was estimated using a simple model (see Section 4.2). Similarly, in intergranular regions, the downflow speed increases from 0.1 to 0.2 km/s with increasing depth in their study. Because of atmospheric seeing, the absorption lines may be affected by blending with blue- and red-shifted profiles, resulting in a smaller Doppler shift. Therefore, the wavelength shift they captured would be smaller than that in our study, indicating that spectral data with high spatial resolution from space enable us to measure the convective structure more accurately. 
A similar tendency is also seen in \citet{Socas-Navarro2011}, which used \textit{Hinode}/SOT spectral data in a quiet region and derived atmospheric parameters including the velocity and geometrical height for each data pixel by NICOLE inversion \citep{Socas-Navarro2015}. Note that the 5-min oscillations were not subtracted before the inversion.
Their atmospheric parameters can be used to derive the velocities as a function of geometrical height when upward and downward regions are considered separately.
The upward regions show deceleration from 0.81 km/s to nearly 0 km/s at geometrical heights ranging from 40 to 160 km, whereas the downward regions show acceleration from 0.67 to 0.86 km/s at heights ranging from 160 to 40 km. As a result, those velocity fields are significantly stronger than those of \citet{Kostik2007}, except for upward regions in higher layers. 
Although the data from the same instrument is used in \citet{Socas-Navarro2011} and our study, the slight difference of the velocity field between them may be due to the difference in techniques and/or the treatment of the 5-min oscillations.
On the other hand, the velocities in our analysis are much smaller than those shown by \citet{Frutiger2000} and \citet{Borrero2002}, who derived their results from inversions with a two-component model of the average spectral profile integrated on the disk center. Our velocities are roughly two times smaller than their values in granules and three times smaller in intergranular lanes. 
Their approach to deriving the velocity field differs significantly from our analysis, which handled each data pixel.
In the future, we will tackle the problem causing these differences between their approach and ours. \\
　Previous works have a large deviation in their reported velocity magnitude due to the possible concerns of the 5-min oscillations and seeing-degradation. 
Thus, we tackled the problem using a times series of stable, seeing free high-resolution data from \textit{Hinode}.
Our analysis provided a clear signature of the deceleration of the convective motion inside granules as a function of height and the acceleration of the convective motion in intergranular regions as the gas descends. 
Moreover, our derived velocities are much larger than those derived in a previous study \citep{Kostik2007} using ground-based telescopes, indicating that spectral data with sub-arcsecond spatial resolution recorded under seeing-free conditions are essential to determining the velocity properties in the photospheric layer more accurately. 
Recently, 1-m-class ground-based telescopes have started to provide sub-arcsecond spatial resolution data and can record various types of spectral lines; these observations greatly extend the range of the geometric height. For example, Fe I lines at 1.5 $\mu$m are helpful for exploring the behavior of convective motion in the deeper layer below the photosphere. Adding to this future analysis, it is also important to compare the bisector results with the velocities obtained using inversion techniques because it may provide some hints on the behavior of convective motion in the photospheric layer. \\
　We now discuss convective stability in the photosphere in light of our result. In Fig. \ref{fig:down_up}, the upward velocity exhibits deceleration as material ascends, whereas the downward velocity shows a trend of acceleration as it descends.
The photospheric layer has been regarded as a convectively stable layer \citep{Stix2004}. 
Larger atmospheric temperature gradients along the vertical direction, compared to the temperature variations of parcels moving adiabatically, can produce convective instability. 
In the photospheric layers, the atmospheric temperature gradient is thought to be smaller than that of gas parcels moving adiabatically, which is considered to indicate convective stability. 
A convectively stable layer weakens the amplitude of the convective velocity; that is, the velocity field is decelerated as material moves upward or downward. 
Our results show decelerating upward convective motion with increasing height in granules, which is compatible with convective stability. On the other hand, the accelerating downward motion with increasing depth in intergranular lanes cannot be explained by convective stability. 
Thus, we need an extra force to break the convective stability in intergranular regions. 
According to radiation MHD simulations (\citealt{Cheung2007} and \citealt{Stein1998}), the following scenario can be proposed to explain these accelerating downflows. After ascending, the material releases its energy by radiative cooling, and its temperature is decreased. 
Consequently, the material becomes denser than its surroundings and is more subject to being pulled down by gravity, leading to acceleration. Another scenario is that a pressure gradient is the driving force causing the acceleration; this possibility is suggested by a previous numerical simulation \citep{Hurlburt1984}. Intergranular lanes have high pressure because material is supplied horizontally from granular regions. The excess pressure would cause acceleration of the submerging materials. 
It is, however, difficult to use our observational result to determine which scenario dominantly controls the acceleration. \\
　Comparisons of the observed temporal behavior with that of numerical simulations are useful to verify the validity of the physics involved in the simulations. Our bisector analysis can provide the temporal evolution of the velocity structure in the vertical direction at photospheric height. Such time series are relevant to the high temporal and spatial resolution data in numerical simulations. As an example, we presented the temporal evolution of a granule that was fragmented during the observation. 
In the central area of granules, the photospheric intensity gradually decreased with time, and a downward flow gradually developed. Downward flow development with decreasing intensity was reported by \citet{Hirzberger2001}, who studied 30 fragmented granules using a ground-based observation. \citet{Berrilli2002} also reported the occurrence of upflows in the area surrounding the decreasing intensity region. 
Our bisector analysis confirms these observational facts regarding the origin of the downflow and its development along the height direction in the photospheric layer. In all four cases, we clearly observed that a downflow appeared in the upper portion of the photosphere and gradually extended toward the bottom of the photosphere. This suggests that radiative cooling works more efficiently on material in the upper portion of the observed photospheric layer, leading to the appearance of downward motion in the upper portion of the central region of granules. After the appearance of downward motion in the upper portion, it took less than 30 s for the downward motion to develop further in the portion located about 120 km below the initial height. Because the observed downflow speed is less than 1 km/s, the initial downward-moving material in the upper layer cannot move to the lower layer on such a short timescale. Rather, we observed a slight difference in the efficiency of radiative cooling along the height direction. \\
　This observational proof of a gradual decrease in the velocity field is consistent with previous numerical simulations reported by \citet{Rast1995} and \citet{Stein1998}. In their simulations, through successive hot gas supply from below the photosphere, the gas at the center of a granule reaches high pressure. It pushes on the surrounding area, and the hot gas prevails horizontally.
The intensity at the periphery becomes higher because of this spreading of hot gas, and the central part loses energy through radiative cooling. Consequently, a dark feature appears at the center, and it splits the granule into smaller cells. We observed this scenario of fragmentation, finding that the convection changes from upward to downward during granular splitting (Fig. \ref{fig:gra_frag}). On the other hand, we did not detect a significant increase in intensity in the region surrounding the granules during fragmentation; this result is supported by the numerical simulations of \citet{Stein1998} and the observational work of \citet{Berrilli2002}. In appearance, only one example shows increasing intensity in newly formed granules, whereas others do not. 

\subsection{5-min oscillations}
　The amplitude of the 5-min oscillations increases from 0.3 to 0.4 km/s with increasing height in the line formation layer (Fig. \ref{fig:rms}). This increasing amplitude is consistent with that in past works. \citet{Deubner1974} found this increasing amplitude of the 5-min oscillations with increasing height by using multiple lines covering heights from the photosphere to the chromosphere. It may be caused by density changes in the photospheric layer; in the solar atmosphere, the density decreases toward the upper atmosphere, and thus the amplitude of sound waves increases as they propagate upward in accordance with the conservation of acoustic energy flux. Using a ground-based telescope, \citet{Kostik2007} also found this increasing tendency using bisector analysis applied to spectral data, including a photospheric line whose maximum formation height is 570 km. The amplitude in their study increases from 0.30 to 0.35 km/s with increasing height from 40 to 160 km. In contrast to convective motion, it seems that oscillation signals derived from ground-based telescope data are not significantly affected by the atmospheric seeing because the 5-min oscillations are a much larger-scale phenomenon than the convection features.
The horizontal spatial scale of the 5-min oscillations ranges from 3 to 4 Mm in the photosphere, corresponding to 4$^{\prime \prime}$ to 6$^{\prime \prime}$ (Fig. \ref{fig:td_sep}). This result suggests that the 5-min oscillation signals can be observed in the velocity field even if the observations do not possess good spatial resolution. 

\section{Summary}
　There is a large deviation in the reported photospheric velocity magnitude in previous works due to the concerns of the 5-min oscillations and seeing-degradation. Thus, we attempted to derive the convective structure by applying bisector analysis to the SP spectral data from \textit{Hinode} with a subsonic filter for the purpose of solving those issues.
Our results show that the convective velocity decreases from 0.65 to 0.40 km/s with increasing height in granular regions, and it increases from 0.30 to 0.50 km/s with increasing depth in intergranular lanes. 
These values are much larger than those in a previous work \citep{Kostik2007} using the same method. 
In granular regions, this result indicates that overshooting materials coming from beneath the photosphere decelerate with increasing height, which is consistent with convective stability in the photosphere. 
Although this stability should also cause descending material to decelerate with increasing depth, our results show the opposite behavior in intergranular lanes. 
Radiatively cooled material tends to become denser and experiences a strong downward gravitational force, resulting in accelerating downward motion. 
The pressure gradient is another possible cause of the acceleration, because the excess pressure in intergranular lanes, where material is supplied horizontally from the granular region, would cause acceleration of the submerging material.
We confirm observationally that a radiative cooling process or pressure gradient greatly affects moving material in the photosphere; it plays a role of a decelerator for upflow in granules and an accelerator for downflow in intergranular lanes.
Bisector analysis is an old method in astronomy, but it is useful when we investigate the dynamical behavior of convective material using a time series of spectral data acquired with high time cadence as far as the dataset have enough time duration and spatial coverage for the removal of 5-min oscillations. Using the spectral profiles from \textit{Hinode}, we provided an example showing the details of dynamical behavior observed in the central portion of fragmenting granules. Downward motion appeared in the upper portion of the photospheric layer and extended to the lower portion in a fairly short timescale in the central portion of fragmenting granules.\\




\acknowledgments
\textit{Hinode} is a Japanese mission developed and launched by ISAS/JAXA, collaborating with NAOJ as
a domestic partner, NASA and STFC (UK) as international partners. Scientic operation of the \textit{Hinode} mission is conducted by the \textit{Hinode} science team organized at ISAS/JAXA. This team mainly consists of scientists from institutes in the partner countries. Support for the post-launch operation is provided by JAXA and NAOJ (Japan), STFC (U.K.), NASA, ESA, and NSC (Norway).
We are grateful to the \textit{Hinode} team for performing the observation on 2014 July 6, which is nicely suited to this analysis. We thank T. Sekii for giving insightful comments on oscillatory motion, D. Dravins for suggesting a reasonable way to estimate an error in the velocity amplitude through the Doppler velocity of the used line in this study, and Dr. D. Brooks for polishing the language of the manuscript. 

\bibliographystyle{apj}
\bibliography{myrefs}

\clearpage





\begin{figure}
\includegraphics[bb=0 0 986 873, width=13cm]{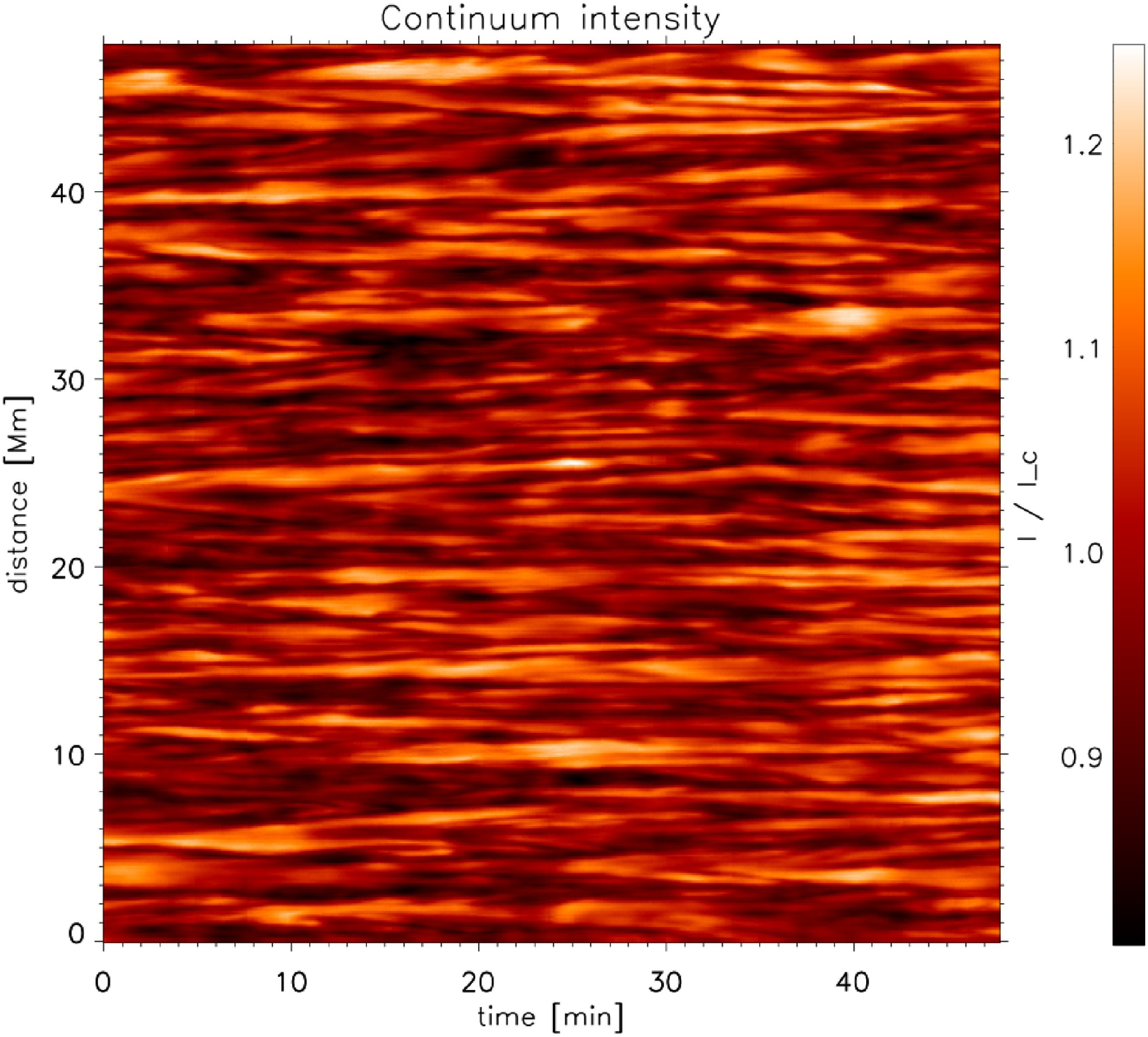}
\caption{Time evolution of the continuum intensity on the slit. The continuum intensity is the intensity averaged over 0.01 nm at 630.1 nm at each pixel and normalized by $I_{c}$, the continuum intensity averaged in the entire FOV. The vertical axis is the slit direction, and the horizontal axis corresponds to time. }
\label{fig:intensity}
\end{figure}

\begin{figure}
\includegraphics[bb=0 0 967 1402, width=11cm]{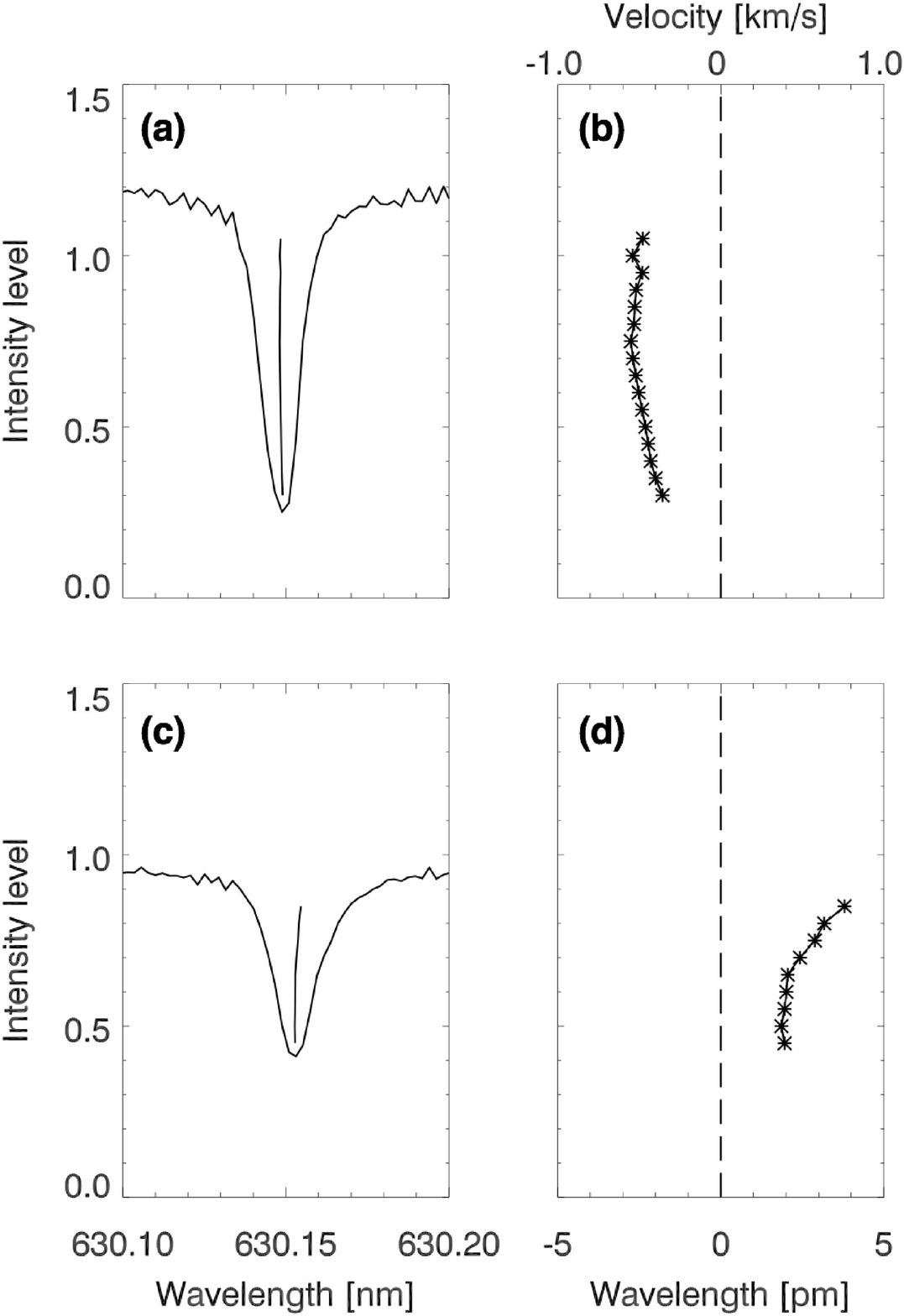}
\caption{(a) Example of an absorption line (Fe I 630.15 nm) in a granule. The solid line in the center of the absorption line is a bisector. (b) Enlarged view of (a). The vertical dashed line is a bisector without the effect of the Doppler shift caused by the moving parcel. The horizontal axis gives the wavelength offset from the absorption line without velocity. (c) Example of an absorption line (Fe I 630.15 nm) in an intergranular lane. (d) Enlarged view of (c).}
\label{fig:bisec}   
\end{figure}

\begin{figure}
\includegraphics[bb=0 0 858 1494, width=11cm]{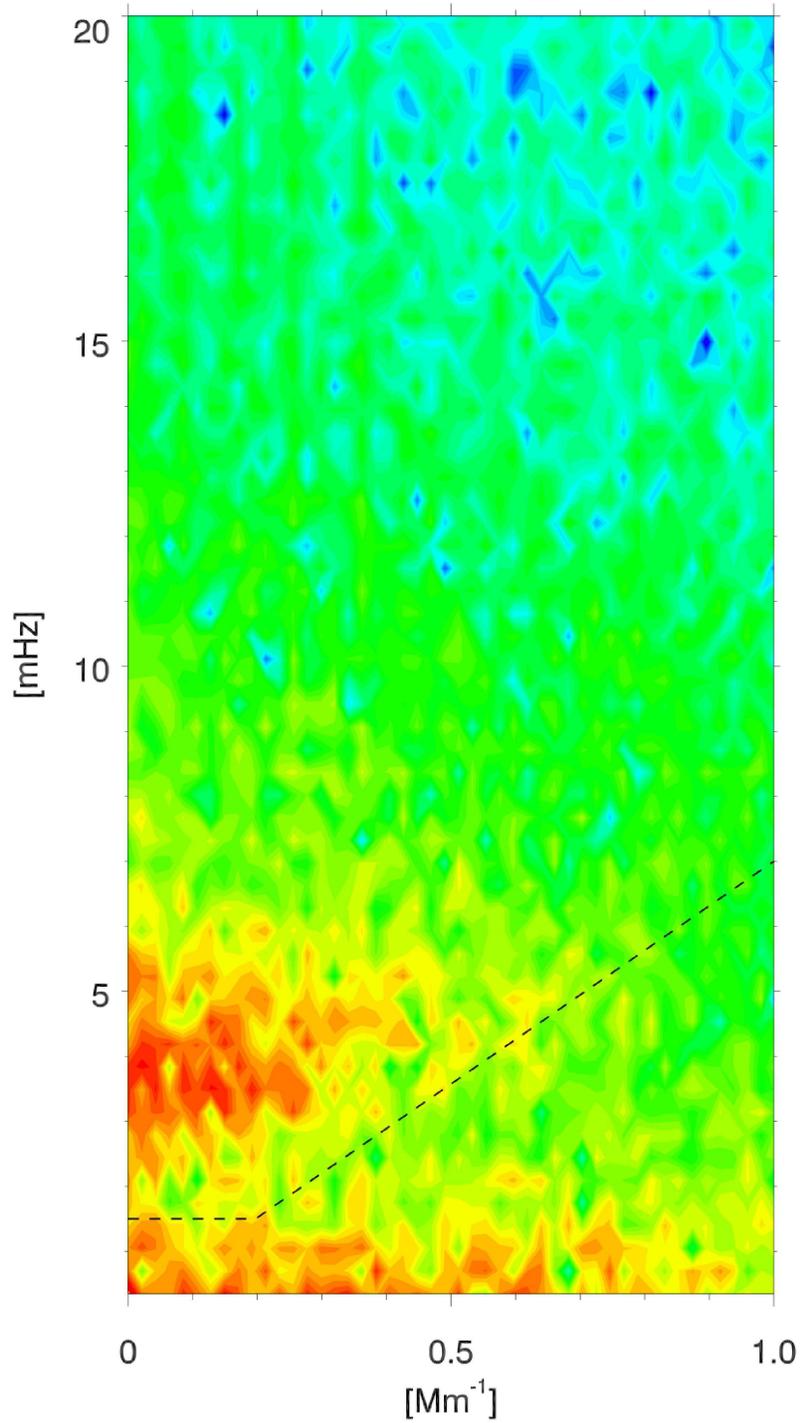}
\caption{$k-\omega$ diagram at an intensity level of $I/I_{c}=0.70$. The black dashed line is a boundary to discriminate between pure granular convective motion in the lower part of the diagram and 5-min oscillations in the higher part. }
    \label{fig:k?_sep}
\end{figure}

\begin{figure}
\includegraphics[bb=0 0 956 1502, width=12cm]{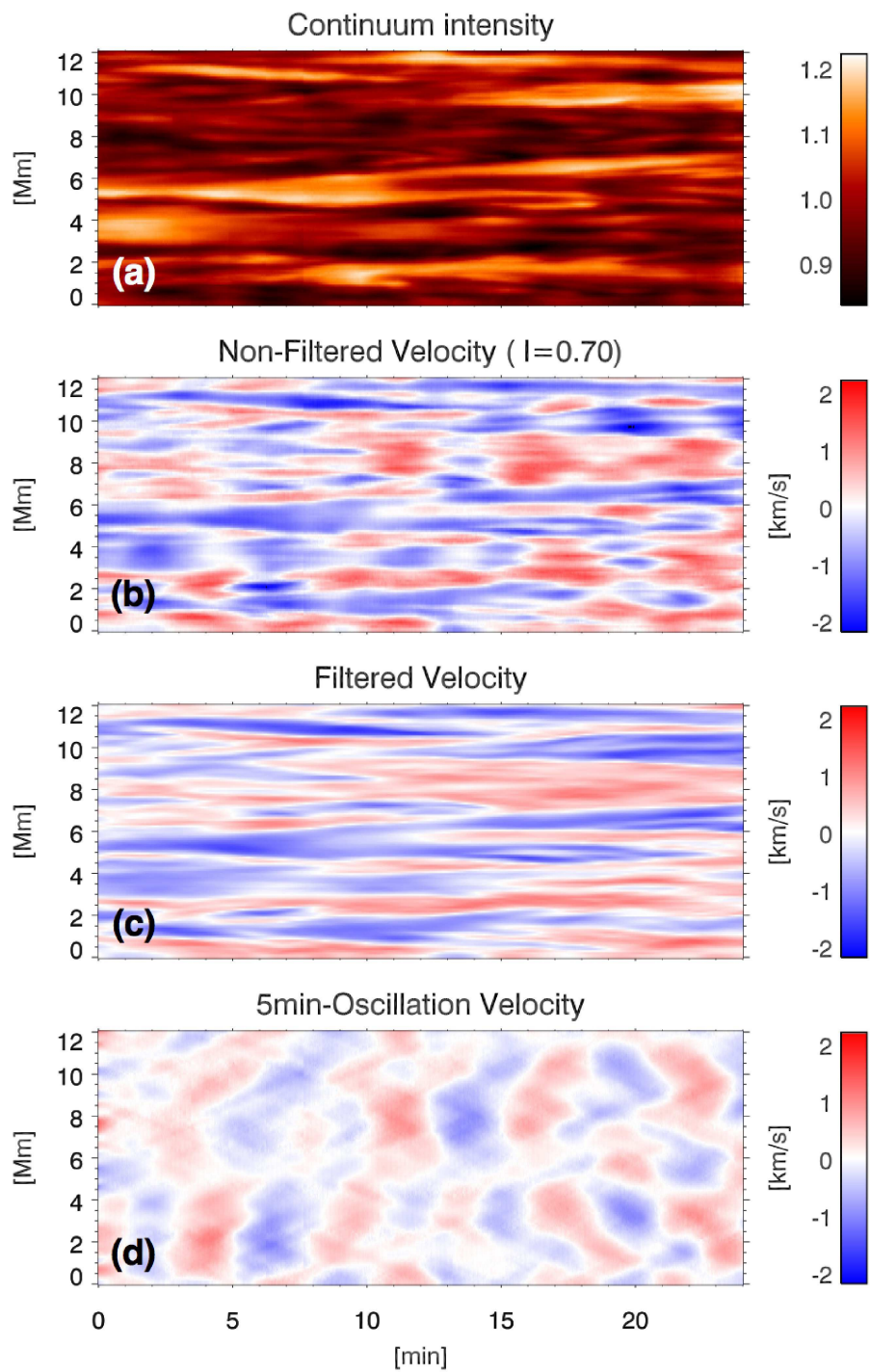}
\caption{Time--distance diagrams of the Doppler velocity for an intensity level of $I/I_{c} = 0.70$ before and after subsonic filtering. (a) Continuum intensity. (b) Time--distance diagram before filtering. (c) Time--distance diagram for convective motion after filtering. (d) Time--distance diagram for 5-min oscillations after filtering. Positive (red) signal is downward, and negative (blue) signal is upward.}
    \label{fig:td_sep}
\end{figure}

\begin{figure}
\includegraphics[bb=0 0 1032 1097,width=13cm]{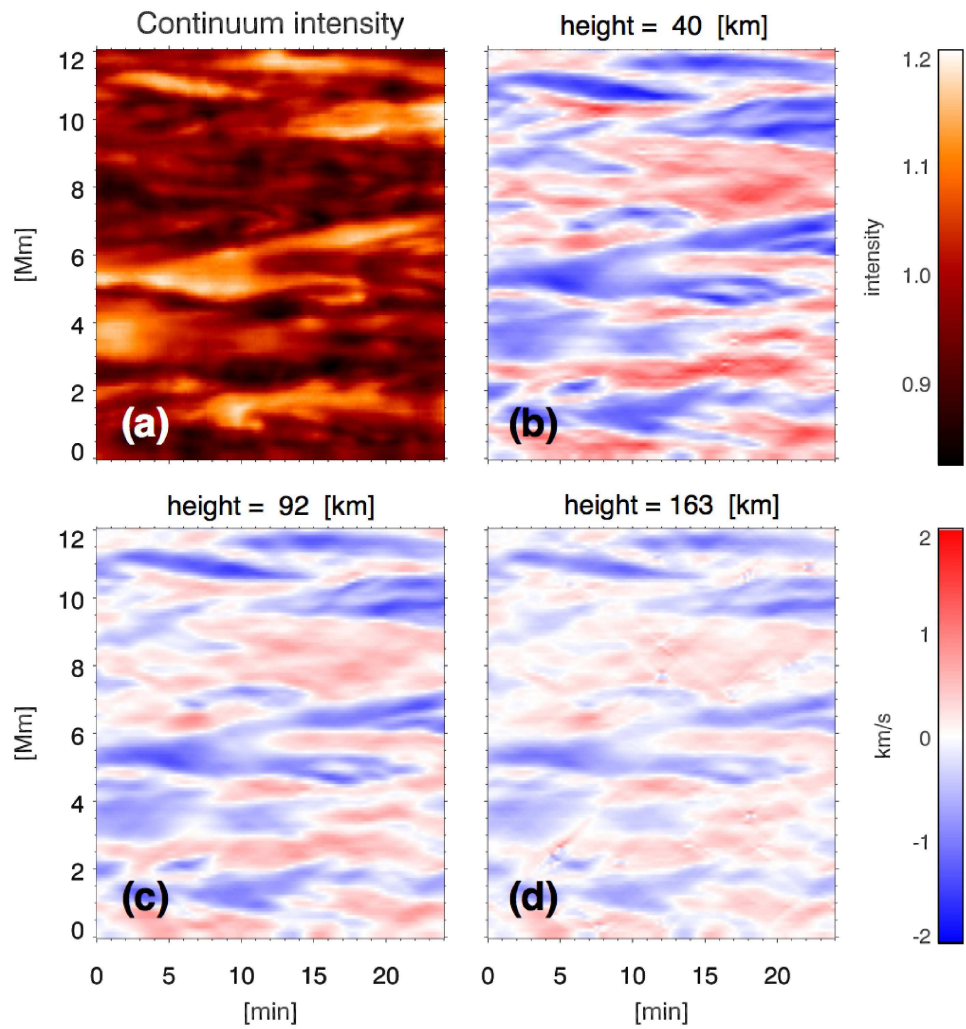}
\caption{Time--distance map of (a) continuum intensity and (b)--(d) convective velocities at intensity levels of 0.75, 0.55, and 0.40, which correspond to heights of 40, 92, and 163 km, respectively. Negative values (blue) are upward, and positive ones (red) are downward.}
    \label{fig:td_cv}
\end{figure}

\begin{figure}
\includegraphics[bb=0 0 949 1461,width=11cm]{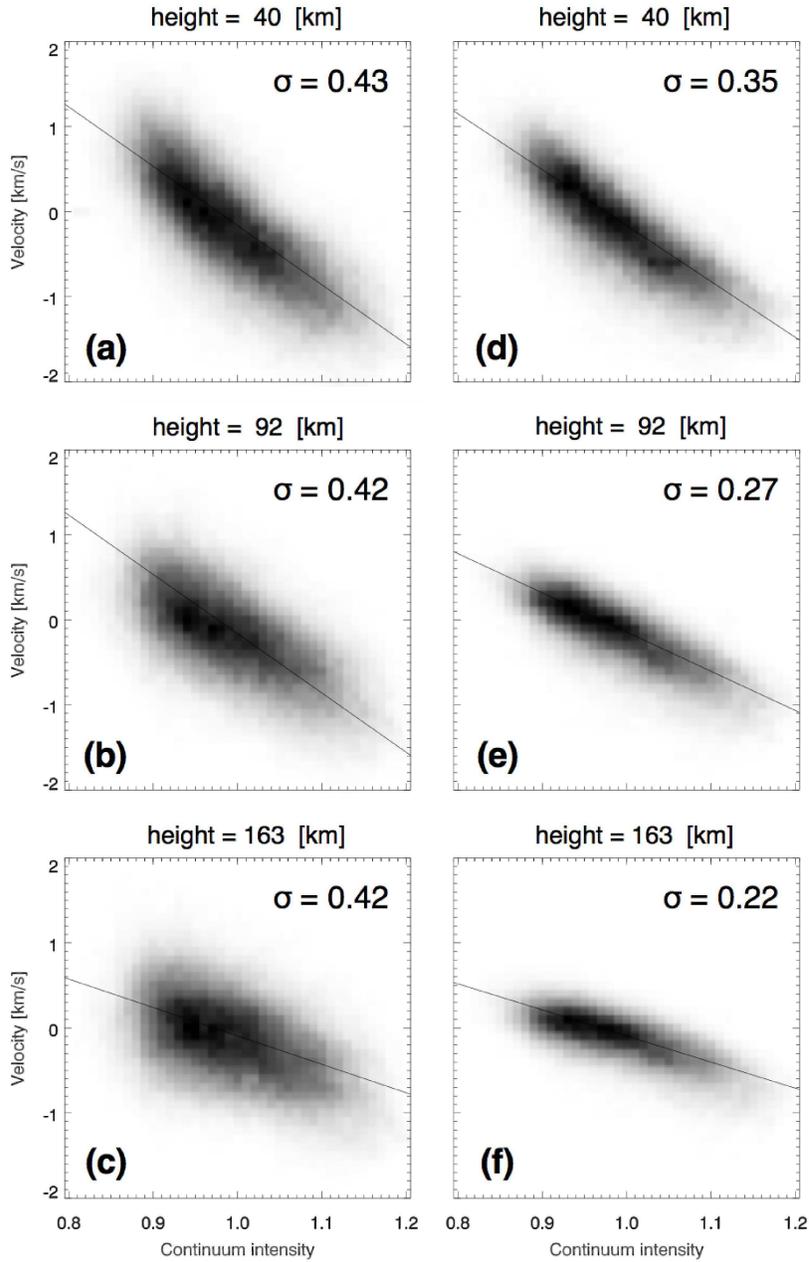}
\caption{Scatter plots between continuum intensity and convective velocity. Note that (a), (b), and (c) are the velocity fields before the filtering process. On the other hand, (d), (e), and (f) are the velocity fields after the filtering process. (a) and (d) are at an intensity level of 0.75, (b) and (e) are at an intensity level of 0.55, and (c) and (f) are at an intensity level of 0.40; these levels correspond to heights of 40, 92, and 163 km, respectively. }
    \label{fig:scp}
\end{figure}


\begin{figure}
\includegraphics[bb=0 0 1014 1199,width=13cm]{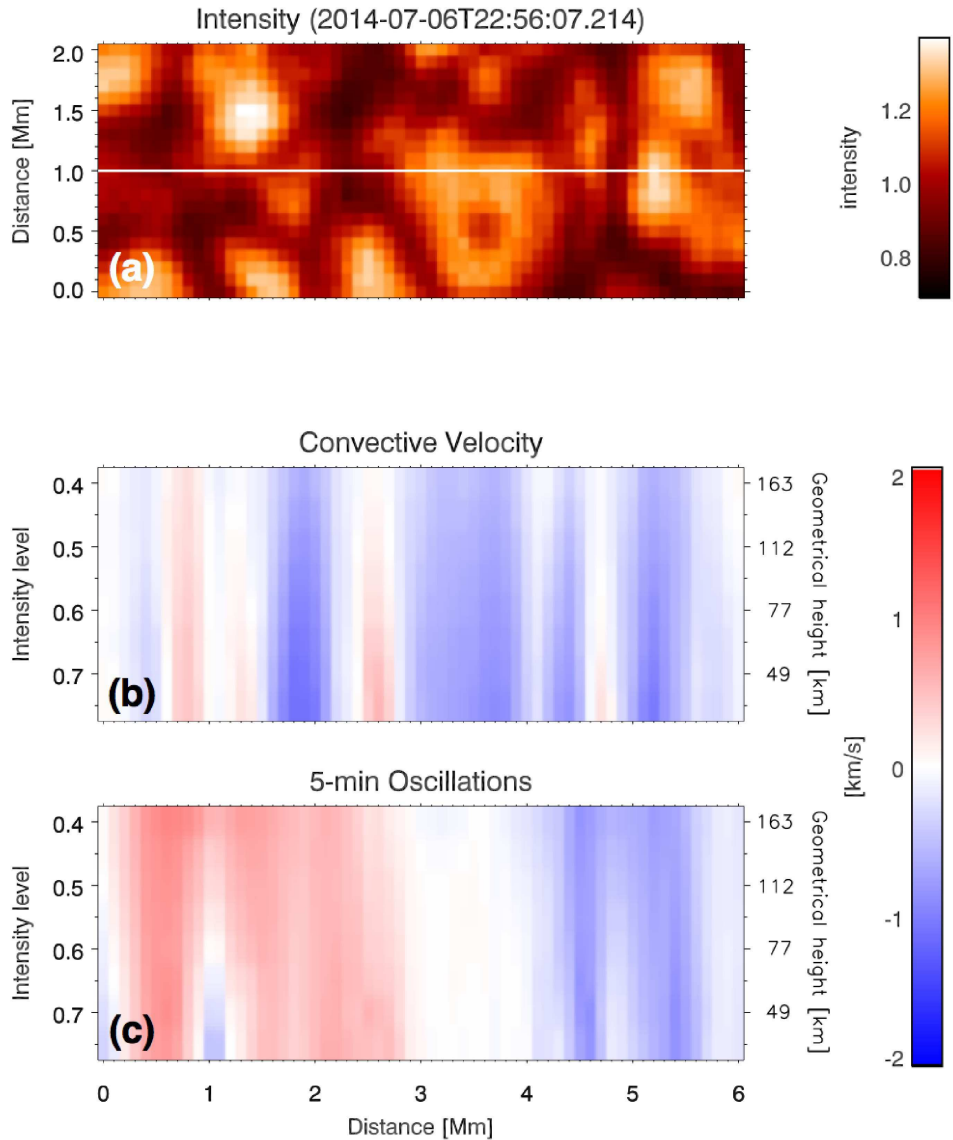}
\caption{Snapshot of the continuum intensity, structure of convective velocity, and 5-min oscillations. (a) Two-dimensional image of the blue continuum intensity. (b), (c) Height-horizontal extent of convective velocity and 5-min oscillations, respectively, along the white line in panel (a). The horizontal axis is along the solar N--S direction.}
    \label{fig:movie}
\end{figure}

\begin{figure}
\includegraphics[bb=0 0 843 1543,width=10cm]{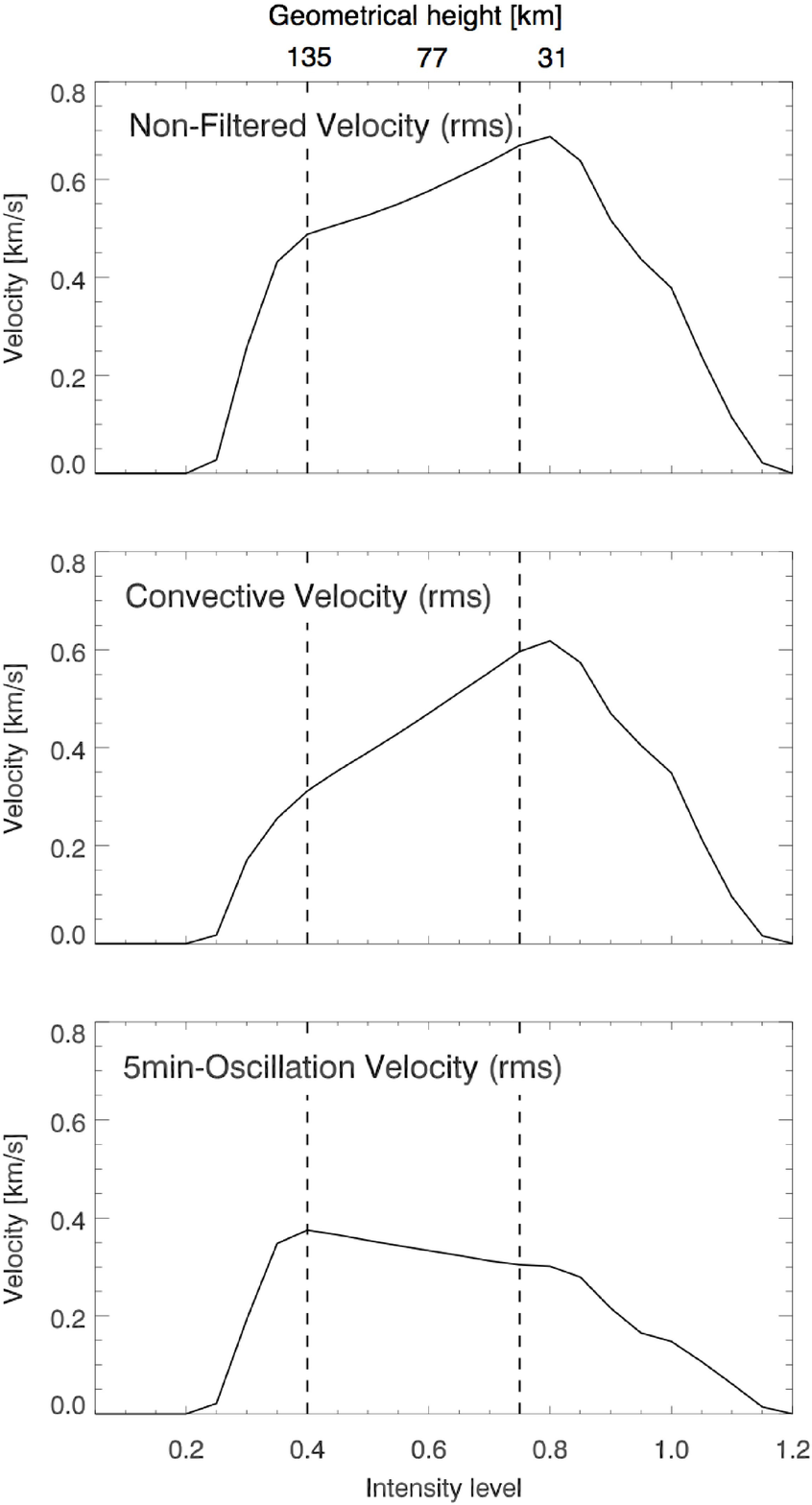}
\caption{RMS values of the Doppler velocity in both granular and intergranular regions as a function of intensity level. From top to bottom: original velocity (unfiltered velocity) before filtering, pure convective velocity derived by filtering (filtered velocity), and velocity of 5-min oscillations after filtering. The intensity levels from 0.40 to 0.75 gives a reliable result, as discussed in the text. The corresponding geometrical heights are given in the frame at the top.}
    \label{fig:rms}
\end{figure}

\begin{figure}
\includegraphics[bb=0 0 918 1524,width=11cm]{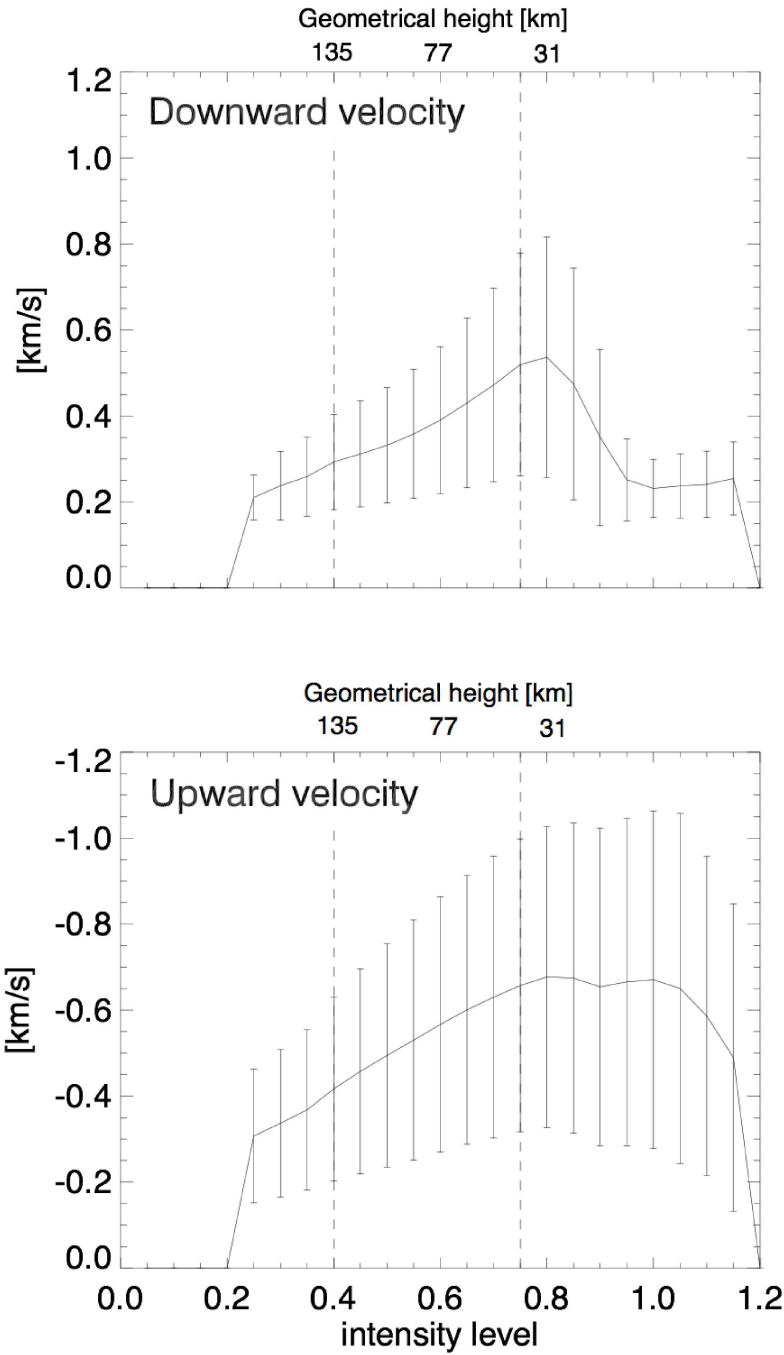}
\caption{Averaged vertical velocity as a function of the intensity level for upflow and downflow regions. \textit{Upper panel}: Downflow regions, defined as regions where the convective speed is higher than 0.18 km/s. \textit{Lower panel}: Upflow regions, defined as regions where the speed is lower than $-$0.18 km/s. Error bars represent the standard deviation.}
    \label{fig:down_up}
\end{figure}

\begin{figure}
\includegraphics[bb=0 0 863 1405, width=11cm]{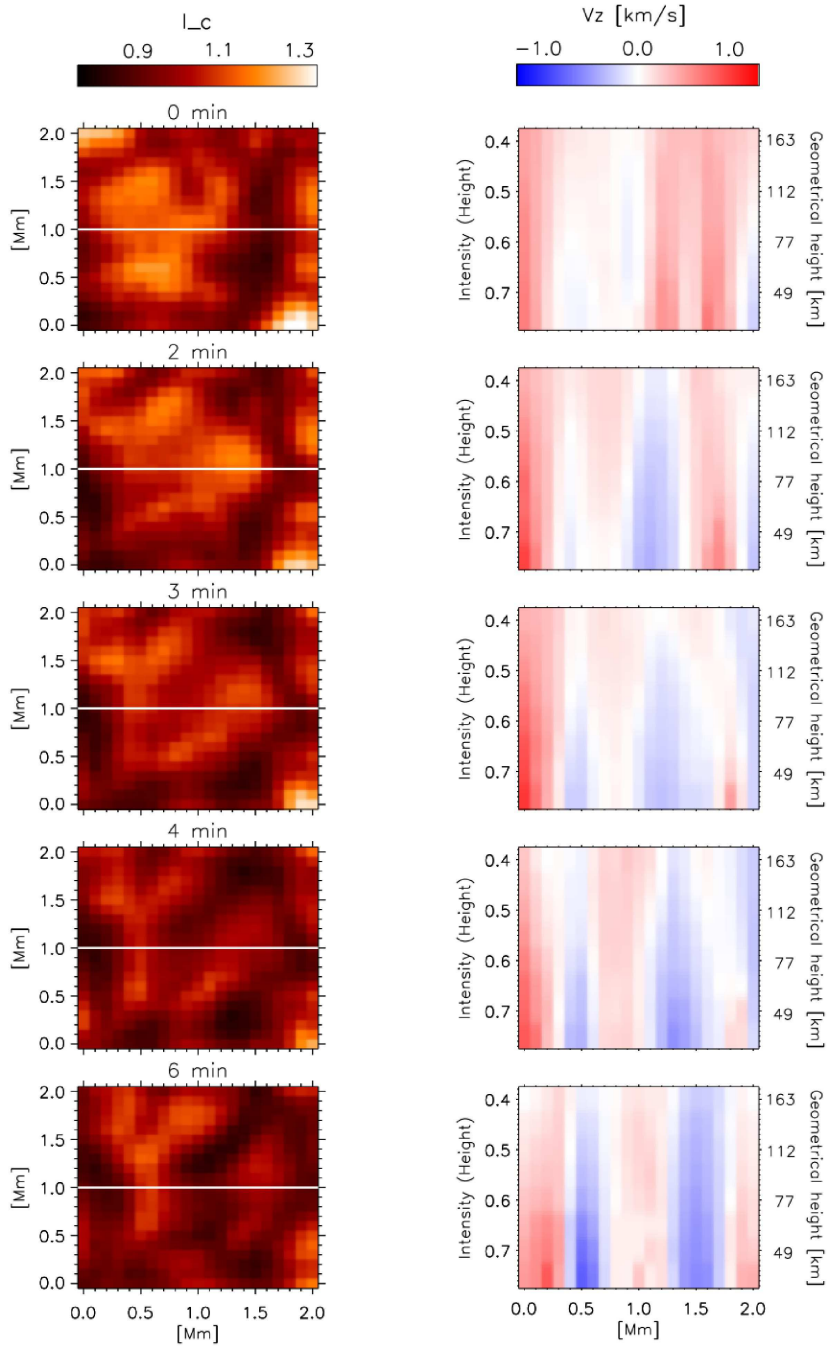}
\caption{Time evolution of velocity structure in a fragmenting granule. \textit{Left}: Blue continuum images from FG observation. The SP slit positions are given by the white lines in each image. The solar N--S direction is along the horizontal axis. \textit{Right}: Convective velocity structure in the slit position. Red shows downward flow, and blue represents upward flow. }
    \label{fig:gra_frag}
\end{figure}

\begin{table}
\begin{center}
\caption{Geometrical height, temperature, and optical depth at 500 nm for each intensity level.}
\begin{tabular}{cccc}
\tableline\tableline
Intensity ($I/I_{0}$) & Temperature [K] & Height [km] & Optical depth ($\tau_{500}$)\\
\tableline
0.75 & 5951 & 40 & 0.55\\
0.70 & 5848  & 49 & 0.44\\
0.65 & 5740  & 62 & 0.38\\
0.60 & 5630  & 77 & 0.32\\
0.55 & 5513  & 92 & 0.25\\
0.50 & 5391 & 112 & 0.19\\
0.45 & 5262  & 135 & 0.14\\
0.40 & 5125  & 163 & 0.10\\
\tableline
\end{tabular}
\end{center}
\label{tab:1}   
\end{table}

\end{document}